\documentclass[%
 aip,
 amsmath,amssymb,
preprint,%
]{revtex4-1}

\usepackage{graphicx}
\usepackage{dcolumn}
\usepackage{bm}
\usepackage[utf8]{inputenc}
\usepackage[T1]{fontenc}
\usepackage{mathptmx}
\usepackage{xcolor}
\usepackage{url}
\usepackage{hyperref}
\usepackage[normalem]{ulem}
\usepackage{subfigure}
\usepackage{float}

\hypersetup{
 colorlinks = true,
 allbordercolors = {white},
 allcolors = {blue},
 }
\usepackage[symbol]{footmisc}

\begin{document}
\title{Pulse propagation in the quiescent environment during direct numerical simulation of Rayleigh-Taylor instability: Solution by Bromwich contour integral method}

\author{Tapan K. Sengupta\footnote{Corresponding author.}}
\email{tksengupta@iitism.ac.in}
\affiliation{Department of Mechanical Engineering, IIT (ISM) Dhanbad, Jharkhand-826 004, India}


\author{Bhavna Joshi}
\affiliation{Department of Mechanical Engineering, IIT (ISM) Dhanbad, Jharkhand-826 004, India}

\author{Prasannabalaji Sundaram}
\affiliation{CERFACS, Toulouse, France}

 


\date{\today}

\begin{abstract}
In: {\it "Three-dimensional direct numerical simulation (DNS) of Rayleigh-Taylor instability (RTI) trigerred by acoustic excitation -- Sengupta et al. {\bf 34},054108 (2022)"} the receptivity of RTI to pressure pulses have been established. It has also been shown that at the onset of  RTI these pulses are one-dimensional and the dissipation of the pressure pulses are governed by a dissipative wave equation. The propagation of these infrasonic to ultrasonic pressure pulses have been studied theoretically and numerically by a high fidelity numerical procedure in the physical plane. The numerical results are consistent with the theoretical analysis and the DNS of RTI noted above. The properties of pulse propagation in a quiescent dissipative ambience have been theoretically obtained from the linearized compressible Navier-Stokes equation, without Stokes' hypothesis.
This analysis is extended here for a special class of excitation, with combination of wavenumbers and circular frequencies for which the phase shift results in an imposed time period is integral multiple of $\pi$, and the signal amplification is by a real factor. Here, the governing partial differential equation (PDE) for the free-field propagation of pulses is solved by the Bromwich contour integral method in the spectral plane. This method, for an input Gaussian pulse excited at a fixed frequency, is the so-called signal problem. Responses for the specific phase shifts integral multiple of $\pi$ can reinforce each other due to the phase coherence. It is shown that these combinations occur at a fixed wavenumber, with higher frequencies attenuated more in such a sequence.       
\end{abstract}

\maketitle

\textbf{Keywords:} Direct numerical simulation; Rayleigh-Taylor instability; perturbation pressure equation; dissipative wave propagation; aeracoustics\\
\section{Introduction}
The canonical wave equation \cite{Whitham74, SenguptaBhum20} has a long history of development starting with that described \cite{DAlembert1750} for the transverse vibration of a string in tension, and for an electromagnetic field in vacuum, \cite{maxwell1865, maxwell54} for an electric ($E$), and associated magnetic fields ($B$). Planar propagation of the dependent variable ($u$) the wave equation is given by, 
\begin{equation} \label{Eq:utt}
    u_{tt} = c^2 u_{xx}
\end{equation}

The comprehensive multi-dimensional analyses of dissipative pressure pulse propagation have been presented recently \cite{Senguptaetal2023, Senguptaetal2024}. Planar propagation with viscous losses has also been reported before \cite{Blackstock2000, Trusler, Morse_Ingard}.
The results presented here consider the governing equation of the perturbation field created by an impulse in the presence of viscous diffusion terms, which were absent in earlier work on acoustics \cite{Feynman65, Feynman69, Whitham74}.   

Here, the governing equation for the fluctuating pressure in a quiescent ambience is obtained by studying the linearized compressible Navier-Stokes equation, with acoustics pressure ten to twelve orders of magnitude smaller than the hydrodynamic pressure \cite{SenguptaBhum20}. The use of global spectral analysis (GSA) with the wavenumber and circular frequency as independent variables provides the dispersion relation \cite{Senguptaetal2023}.  

The GSA \cite{Sengupta2013} distinguishes the wave-like propagation of pressure pulse from the pure diffusion of the perturbation field. The metrics used in GSA involve the amplification factor and the group velocity, as explained originally \cite{Rayleigh, Lighthill, Brillouin} for acoustics, fluid flow, and electromagnetic wave propagation, respectively. Here, the GSA considers perturbations in flows and acoustics simultaneously which shows the phase speed is not the same as the group velocity due to its dispersive nature \cite{Acoustic_POF18}. Compared to the classical wave equation, viscous stresses in the present governing equation make the system dispersive. The GSA  \cite{Senguptaetal2023, Senguptaetal2024} identifies a cut-off wavenumber ($k_c$), above which the characteristics of the governing equation change from an attenuated wave equation to a strictly diffusive equation. 

The governing equation for the perturbation pressure in a quiescent dissipative medium is given \cite{Senguptaetal2023} by,

\begin{equation} \label{Eq:12}
\frac{\partial^2 p'}{\partial t^2}  = c^2\nabla^2 p' + \nu_l \frac{\partial }{\partial t} \nabla^2  p'
\end{equation}

\noindent with the generalized kinematic viscosity defined as, $\nu_l = \frac{\lambda + 2 \mu}{\bar \rho}$, that accounts for the viscous losses during propagation of the pulse; either as waves or as diffusive disturbances depending on the wavenumber. Here, $\lambda$ is the second coefficient of viscosity and $\mu$ is the dynamic viscosity. Derivation of Eq.~\eqref{Eq:12} does not require the Stokes' hypothesis  \cite{stokes1851}, relating the first and second coefficients of viscosity combining to yield the bulk viscosity defined as $\mu_b = \lambda + \frac{2}{3} \mu$. The role of $\mu_b$ during the propagation of the waves is described during the onset of the RTI, while solving the compressible Navier-Stokes equation\cite{Acoustic_POF14, Acoustic_POF13}.   

The reformulation of pulse propagation reported with $\nu_l$ as a parameter \cite{Senguptaetal2024}, provides the amplification factor over a chosen time scale, the phase speed and the group velocity. This parameter assists in understanding the properties better, as compared to the diffusion number as a fixed parameter \cite{Senguptaetal2023}. Also, a numerical solution of the planar pulse propagation in the free field in the quiescent ambience is provided for an initial Gaussian pulse applied impulsively, by using high accuracy compact scheme with fourth-order Runge-Kutta time integration scheme \cite{Senguptaetal2024}. The numerical solution in the physical plane is accurate and obtained efficiently, without a need to find the spectral weight of each frequency associated with the application of the initial impulse. But this also requires two initial conditions for the physical plane solution.

Here, the signal problem associated with pulse propagation is studied, where the input and its response is assumed to be at the same frequency. Thus, the search is for the frequency response of the dynamical system \cite{POF1994, IFTT, Sengupta21} for flow transition. The signal problem is solved by performing the Bromwich contour integral \cite{Papoulis, VanderPol_Bremmer, IFTT, Sengupta21} in the spectral plane. This is reported here for the first time for pressure pulse propagation, whose governing equation in multi-dimension is given by Eq. ~\eqref{Eq:12}.

The one-dimensional (1D) planar propagation of the perturbation field Eq.~\eqref{Eq:12} simplifies to, 

\begin{equation} \label{Eq:13}
\frac{\partial^2 p'}{\partial t^2}  - c^2\frac{\partial^2 p'}{\partial x^2} - \nu_l \frac{\partial^3  p' }{\partial t\partial x^2}  = 0
\end{equation}

The core of the present analysis is to classify the wavy and non-wavy nature of the solution of Eq.~ \eqref{Eq:13}. The non-wavy nature of the response is based on the necessary condition of vanishing the imaginary part of the amplification factor in GSA. It has been shown\cite{Senguptaetal2023, Senguptaetal2024} that there exists a critical wavenumber ($k_c$) above which the attenuated wavy solution transforms into a purely diffusive solution for the propagation of the pulse in a quiescent ambience.  

The pressure pulse created in a quiescent ambience was treated for the first time \cite{Senguptaetal2023} showing the properties, with the highlight being the presence of $k_c$, above which the wavenumber components will be rapidly attenuated by diffusion. The present research is a complement of the studies recently \cite{Senguptaetal2024}, where we focus on all wavenumbers, including the sub-critical ones ($k<k_c$). The in-depth probe here is by displaying pulse propagation of signals that span across infrasonic to ultrasonic frequency ranges. 

The paper is formatted in the following manner. A brief description of the GSA and Bromwich contour integral is provided in the next section. In section III, special sub-critical solutions with phase coherence are introduced. Theoretical analysis identifies these special sub-critical points with phase shifts given by integral multiple of $\pi$ in this section. In section IV, these specific cases of pulse propagation for a fixed generalized kinematic viscosity are presented using the Bromwich contour integral method. Noted propagation patterns are explained with the help of the physical properties obtained by GSA. The paper closes with a summary and conclusion in section V. 

\section{GSA and Bromwich contour integral method}
Here, the GSA is used for the governing equation. \cite{Blackstock2000, Senguptaetal2023} for the small amplitude perturbation fields in 1D for the physical properties. The GSA is performed for the 1D perturbation field Eq.~\eqref{Eq:13}. This arises as the linearized response for free-field propagation in flows and acoustics caused by a pulse. The response of such an impulse is presented in the spectral plane by the bilateral Fourier-Laplace transform in \cite{Papoulis, VanderPol_Bremmer, IFTT} for the pressure as, 
\begin{equation} \label{Eq:14}
p'(x,t) = \int \int \hat p (k,\omega)e^{i(kx-\omega t)} dk d\omega 
\end{equation}

Using Eq.~\eqref{Eq:14} yields the physical dispersion relation as, 
\begin{equation} \label{Eq:15}
\omega ^2  + i\nu_l k^2 \omega - c^2 k^2 = 0 
\end{equation}

The presence of the imaginary term involving $\nu_l$ implies that both the modes are attenuated due to viscous action. In the GSA, instead of Eq.~\eqref{Eq:14}, one uses the hybrid form as, 
\begin{equation} \label{Eq:14GSA}
p'(x,t) = \int_{Br} \hat p (k,t)e^{ikx} dk  
\end{equation}

As explained  \cite{VanderPol_Bremmer, IFTT, Sengupta21}, the integral on the right-hand side is performed along the Bromwich contour, with $Br$ defined in its strip of convergence. For the GSA of Eq.~\eqref{Eq:13}, a length-scale ($L_s$) and a time-scale ($\tau_s$) are introduced, where $L_s$ can be set up in an arbitrary manner, while we have a special reason in choosing $\tau_s$. For an excitation that is localized in space, all wavenumbers are excited with equal weight. According to Eq.~\eqref{Eq:15}, each wavenumber is associated with a pair of circular frequencies ($\omega_{1,2}$). Thus, for all the wavenumbers excited, one would note infinite frequencies excited. For any frequency, we ascribe $\tau_s$ as the corresponding time period.
One defines the physical amplification factor with respect to $\tau_s$, as the ratio of the amplitudes at an interval of $\tau_s$, with the dependence given by,
\begin{equation} \label{Eq:Amp_Fac1}
G_{1,2} (k;\tau_s) = \frac{\hat p(k,t+\tau_s)}{\hat p(k,t)}
\end{equation}

For the response field, the initial conditions help in defining the Bromwich contour in the wavenumber plane as, 
\begin{equation} \label{Eq:Input}
p'(x,0) = \int_{Br} p_0 (k) e^{ikx} dk 
\end{equation}

\noindent where the origin of the physical plane is located at the site of the impulsive pulse. The corresponding Bromwich contour in the complex circular frequency plane is dictated by the principle of causality\cite{IFTT}. The exact positions of the Bromwich contours are determined by the group velocity indicating the direction of propagation of disturbances \cite{IFTT}. 

Results in non-dimensional form are presented using nondimensional wavenumber and time as $kL_s$, and $N_\tau = c\tau_s/ L_s$, respectively. Here, for each time period $\tau_s$, numerous time steps are used. Using the GSA, one obtains the physical amplification factor reported \cite{Acoustic_POF18} as,
\begin{equation} \label{Eq:Amp_Fac2}
G_{1,2} (k;\tau_s) = e^{-i\omega_{1,2} \tau_s}
\end{equation}
\noindent where $\omega_{1,2}$ are obtained from Eq.~\eqref{Eq:15} as, 
\begin{equation}\label{Eq:16}
\omega_{1,2} = \frac{-i \nu_l k^2}{2} \pm kcf
\end{equation}
\noindent with, $f = \sqrt{1-\left(\frac{\nu_l k}{2c} \right)^2}$, as the factor defining the deviation of the dispersion relation from its non-dissipative, isentropic counterpart of the classical wave equation, for which $\nu_l = 0$ and $f\equiv 1$. The exponents of the amplification factors in Eq.~\eqref{Eq:Amp_Fac2}, indicate corresponding phase shifts over $\tau_s$ given by, 
\begin{equation}\label{Eq:19}
\beta_{1,2} = \pm kcf~\tau_s
\end{equation}
The phase speed and the phase shift are related for the nondimensional phase speeds of the dissipative perturbation equation as, 
\begin{equation} \label{Eq:21}
    \frac{c_{ph 1,2}}{c} = \frac{\beta_{1,2}}{kc~\tau_s} = \pm f 
\end{equation}

Corresponding components of the group velocity ($v_{g 1,2}$) of the perturbation field are given as, 
\begin{equation} \label{Eq:22}
    v_{g 1,2} = \frac{d \omega_{1,2}}{dk} = \pm cf \mp \frac{\left(k \nu_l \right)^2}{4fc} - i \nu_l k
\end{equation}
The real and imaginary parts of the first physical amplification factor are written as,
\begin{equation} \label{Eq:G1}
(G_{1})_{real} = e^{-\frac{k^2 \nu_l}{2}\tau_s} \cos (kfc\tau_s), \;\; \;\; (G_{1})_{imag} = -e^{-\frac{k^2 \nu_l}{2} \tau_s} \sin (kfc\tau_s)
\end{equation}

For sub-critical range of wavenumbers ($k < k_c$), discrete combinations of $kL_s$ and $N_{\tau}$ are reported here, for which the imaginary parts of the amplification factors vanish. However, it is shown here that unlike the purely diffusive properties for super-critical wavenumbers ($k \ge k_c$), for these cases there are discrete loci in $(N_\tau, kL_s)$-plane where the imaginary part of the physical amplification factor vanish due to the phase shift becoming integral multiple of $\pi$.  
In the present research, the focus is not only on the properties for these sub-critical wavenumbers, but also on the pulse propagation and how those depend upon the special properties for these cases of $k \le k_c$. For such wavenumbers, a search is made in an extended ($kL_s$, $N_\tau$)-plane for which the amplification factors are strictly real because the phase shifts are integral multiple of $\pi$. These special sub-critical cases are demonstrated here. 

For the case considered \cite{Acoustic_POF21}, with measurements reported for the attenuated acoustic signal, one notes $c = 343.11 {\rm m/s}$ for air at an ambient temperature of $20^o$C and $\nu_l = 0.1443 {\rm m^2/s}$, which fixes $k_c= 4755.568 {\rm m}^{-1}$. The maximum wavenumber ($k_{max}$) is chosen as four times the value of $k_c$. This helps us choose the smallest resolved length-scale to be given by $L_s = 3.30485 \times 10^{-4}$m, which has been  used in the following. The amplification factors for sub-critical cases are given by,
\begin{equation} \label{Eq:G12_dim}
G_{1,2} (k;\tau_s) = e^{-\nu_l k^2 \tau_s/2} e^{\mp i kL_s N_{\tau} \sqrt{1- (k/k_c)^2}}
\end{equation}
The corresponding phase speeds are given in the dimensional form as, 
\begin{equation} \label{Eq:c12_dim}
    c_{ph 1,2} = \pm c \sqrt{1- (k/k_c)^2} 
\end{equation}
The corresponding group velocity components are obtained from the real part of, 
\begin{equation} \label{Eq:v_g12_dim}
    v_{g 1,2} = \pm c\sqrt{1- (k/k_c)^2} \mp \frac{\left(k \nu_l \right)^2}{4c\sqrt{1- (k/k_c)^2}} - i\nu_l k
\end{equation}
The phase speed and the group velocity are not functions of $N_\tau$. The expressions given in Eqs.~\eqref{Eq:G12_dim} to \eqref{Eq:v_g12_dim} help one to plot the solution properties with $\nu_l$ as a constant.

It is shown \cite{Senguptaetal2023} that for $k > k_c$, the exponent in the second factor becomes purely real, thereby augmenting the first attenuating factor, and $|G_1|$ shows a visible discontinuous jump across the $k_c L_s$-line. Thus, the $k_c L_s$-line demarcates the mathematical characteristics, across which the attenuated wavy solution given by the hyperbolic PDE transforms to the diffusive parabolic PDE. It is also noted for the two modes, the second exponent in Eq.~\eqref{Eq:G12_dim} also changes sign. Thus for the signal created in a quiescent ambience, the two modes are not same above $k \ge k_c$, due to the viscous term with a mixed derivatives of order three. However, for $k < k_c$, the properties are perfectly identical for $|G_1|$ and $|G_2|$. 

Detailed description of the numerical solutions for some specific cases are provided next. The physical scenario consists of a Gaussian wave-packet given by the initial condition,
\begin{equation}\label{Eq:initial}
    p'(x,0)    = e^{(-\alpha(x-x_0)^2)}sin(k_0(x-x_0))
\end{equation} 
\noindent where $x_0 = 0$ and $\alpha =500$. To solve Eq.~ \eqref{Eq:13}, a second initial condition needed for the solution in the physical plane was given by \cite{Senguptaetal2024},
\begin{equation}\label{Eq:initial_derivative}
    \frac{\partial p'}{\partial t} (x,0) = 0 \end{equation} 
For the time-impulsive start of the Gaussian pulse, we note that all the circular frequencies will be excited with equal emphasis. 

\section{Special sub-critical solutions with phase coherence}
\begin{figure*}
\centering
\includegraphics[width=0.9\textwidth]{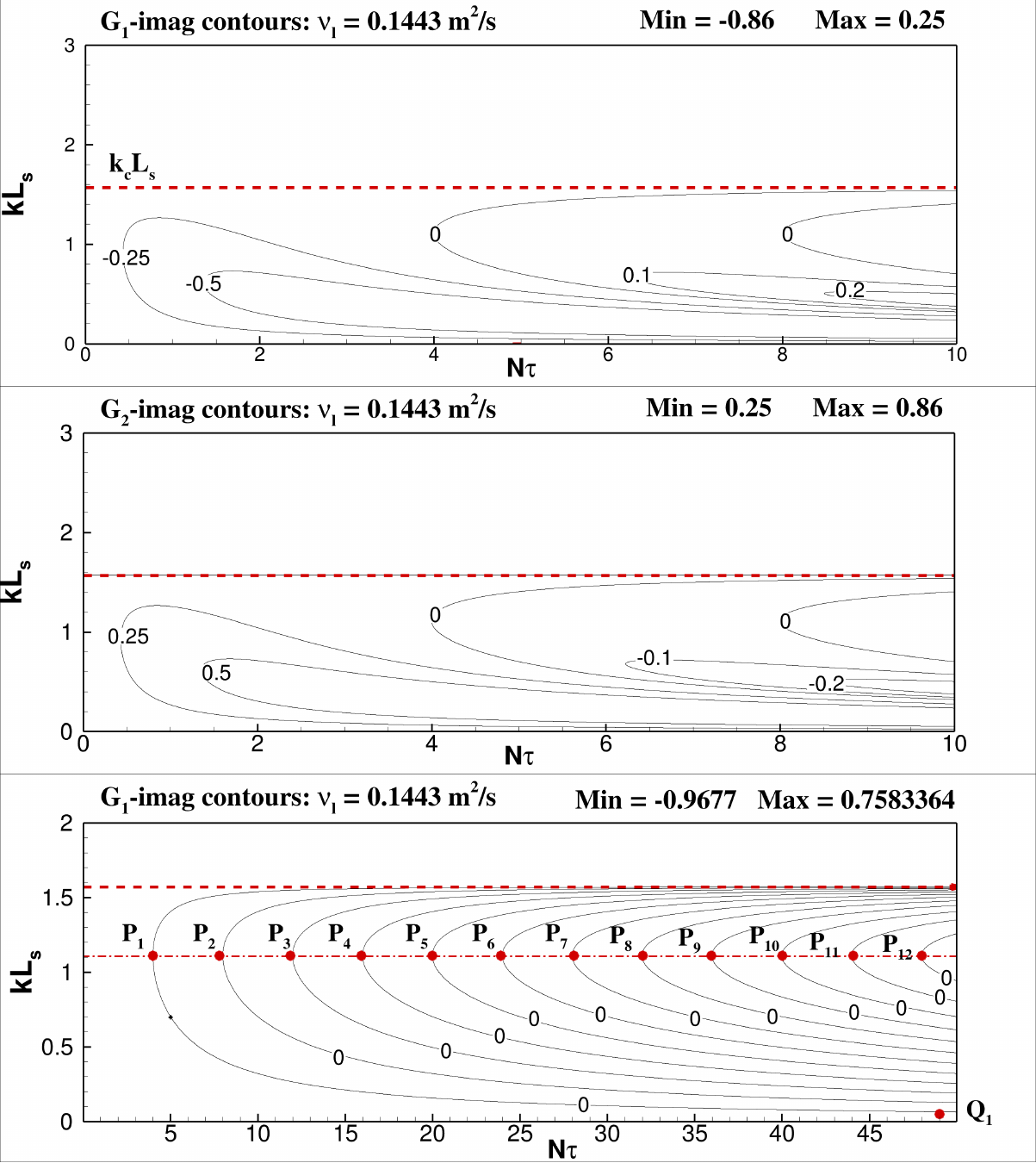}
\caption{The frames show the imaginary part of the amplification factor for the first and second modes of the amplification factor in the top and middle frame, respectively for $\nu_l = 0.1443m^2/s$ for $N_\tau \le 10$. The bottom frame shows the imaginary part of $G_1$ in an extended range of $N_\tau$ up to 50.}
\label{multi}
\end{figure*}

The dispersion relation of the disturbance field is provided by the complex conjugate relations given in Eq.~\eqref{Eq:16} for the left- and right-running wave solutions for sub-critical cases ($k \le k_c$). However, for each of these right-running and left-running waves, it is possible to have parameter combinations for $kL_s$ and $N_\tau$, for which the imaginary part of the amplification factors will be absent. The imaginary part of $G_1$ will be absent in Eq.~\eqref{Eq:G1}, when the phase shift for the time interval of $\tau_s$ for the waves is an integral multiple of $\pi$, i.e. a situation for the first mode, with $\beta_1 = m \pi$, for all integral values of $m$. This implies that,  
\begin{equation} \label{Eq:beta1}
kcf \tau_s = m\pi {\hspace{5mm}} {\rm for}~m = 1,2, .... \infty
\end{equation}
For the general case, calling this wavenumber as $k_m$, the above condition is,

\begin{equation} \label{Eq:beta2}
k_m \sqrt{1 - (k_m / k_c)^2} = \frac{m\pi}{c\tau_s} 
\end{equation}
\noindent where the cut-off wavenumber is defined as before by, $k_c = 2c/ \nu_l$. For $k \ge k_c$, $\omega_{1,2}$ is strictly imaginary, and therefore $G_{1,2} (k; \tau_s)$ is strictly real, as one notes for parabolic PDEs. For $k < k_c$, the circular frequency and the physical amplification factor are in general complex, and the spatiotemporal dynamics display an attenuated wavy nature, i.e. the governing equation is given by a hyperbolic PDE, with $G_{1,2}$ as complex conjugates. However, for $k_m \le k_c$, the amplification factors are strictly real, yet the phase speed and group velocity are non-zero, for integral values of $m$. This interesting possibility has not been identified before, and is the central concern of the present research. 

With the help of $L_s$ and $N_\tau$, the condition given in Eq.~\eqref{Eq:beta2} is written alternately as,
\begin{equation} \label{Eq:locus}
N_\tau k_m L_s \sqrt{1 - (k_m / k_c)^2} = {m\pi} 
\end{equation}
In Fig. \ref{multi}, the imaginary contours of $G_1$ and $G_2$ are plotted in $(N_\tau, kL_s)$-plane for the indicated ranges, with the minimum and maximum values of the contours identified for $\nu_l= 0.1443 m^2/s$. The non-dimensional critical wavenumber ($k_c L_s$) can be identified by the dashed (red) line. One interesting aspect is the presence of multiple contours along which the imaginary part of the amplification rates is zero. 

One can plot the contours given by Eq.~\eqref{Eq:locus} for different values of $m$ in the $(N_\tau, kL_s)$-plane, as shown in the bottom frame of Fig. \ref{multi}. Along these special contours, the amplification factors simplify to,
\begin{equation} \label{Eq:G12_multi}
G_{1} (k,\tau_s) = e^{-\nu_l k^2 \tau_s/2} e^{-im\pi} \;\;\;\; {\rm and} \;\;\;\; G_2 (k,\tau_s) = e^{-\nu_l k^2 \tau_s/2} e^{im\pi}
\end{equation}
This shows the amplification factors as strictly real, as shown in Fig. \ref{multi}, for different integral values of $m$. The loci of points in the $(N_\tau, kL_s)$-plane for the sub-critical wavenumbers ($k < k_c$) are shown in the bottom frame of this figure, that follows Eq.~  \eqref{Eq:locus}. It implies that the solution represent waves with non-zero phase speed and group velocity, yet the amplification factors is strictly diffusive for  such loci with integral values of $m$. Along these contours of Fig. \ref{multi} shown in the bottom frame, the phase shift corresponding to $\beta= \pm m\pi$ will have phase speed given by, $\beta_{1,2} =k c_{1,2} \tau_s$, i.e. $c_{1,2} = \pm (m\pi)/(k\tau_s)$. 

The presence of larger number of the parameter combinations for which the imaginary part of the amplification factor disappears can be noted for extended range of $N_\tau$  displayed in the bottom frame of Fig. \ref{multi}, with $N_\tau$ up to 50. In the bottom frame, one can identify twelve such contours for combinations of $N_\tau$ and $kL_s$. 

One can observe from the results given by Eq.~\eqref{Eq:G12_multi} and Fig. \ref{multi} the following. Along any of the zero imaginary contours of $G_{1,2}$ in the figure, the phase shift remains the same, despite $kL_s$ and $N_\tau$ keep changing along the curve, and any interaction among points lying on such curves will mutually reinforce each other. This will also be true for contributions coming from different zero contours for different values of $m$ due to phase coherence. Secondly, the attenuation will be least for the tip of each of these curves, which have been labeled as $P_1$, $P_2$ ...... $P_{12}$ in the bottom frame of Fig. \ref{multi}. In the following sub-section, the exact locations and their wave properties are described.

\subsection{Sub-critical attenuated wave solutions with phase coherence}

In the bottom frame of Fig. \ref{multi}, the special points marked as $P_1, P_2, P_3 ......P_{12}$, all belong to the zero-contours of imaginary part of $G_{1,2}$, and thus have same phase relationships, and are either in phase or in anti-phase, depending upon the values of $m$. It is noted that these points have the same value of $kL_s$, but attenuate more and more, as $m$ increases. Thus, for an applied impulse, the response will be dominated by these special points, with $P_1$ making the major contribution for the impulse that theoretically excite all possible wavenumbers and circular frequencies.

These special points at the tip of the zero-contours for the successive phase shift cases given by Eq.  \eqref{Eq:locus} are for different values of $m$ as the subscript. If we denote, $x= N_\tau$ and $y= k_m L_s$, then Eq.~\eqref{Eq:locus} is rewritten as,
$$ xy \sqrt{(1 - y/y_c)^2} = m\pi$$
\noindent where $y_c = k_c L_s$. It is evident from Fig. \ref{multi} that these discrete tip-points are located for the same value of $y = y^m_*$, and satisfy the following condition: $\frac{dx}{dy} = 0$. Upon satisfaction of these conditions, one obtains the following values for the special points given by,
\begin{equation} \label{Eq:x*y*}
x^m_* = 2m\pi/ (k_c L_s) \;\;\; {\rm and} \;\; y^m_*=  (k_c L_s)/ \sqrt{2} 
\end{equation}
Corresponding phase speed and group velocity are obtained from Eqs.~ \eqref{Eq:c12_dim} and \eqref{Eq:v_g12_dim} as 
\begin{equation} \label{Eq:cph}
c_{ph 1,2*} = \pm c \sqrt{1 -(y^m_*)^2/y_c^2}=\pm c/\sqrt{2} \;\;\; {\rm and} \;\; v_{g 1,2*}= \pm c \sqrt{2} 
\end{equation}
These are the central results  presented in this subsection. The response field due to an impulse is dispersive\cite{Senguptaetal2023}. The analysis reasons that the response will be dominated by some special points, which have coherent phase relation and belong to the zero-contour lines of the imaginary part of $G_{1,2}$. These points lying on the discrete curves have phase shift given by $m\pi$ in the ($N_\tau, kL_s$)-plane, and which reinforce and nullify each others' effects. These points $P_1$, $P_2$, .... $P_{12}$.... correspond to the least value of $N_\tau$ for each curve, will dominate, with all of them having identical phase speed and group velocity, given in Eq. \eqref{Eq:cph}. 
Using Eq. \eqref{Eq:x*y*} in \eqref{Eq:G12_multi}, one gets the amplification factors for these special points as,
\begin{equation} \label{Eq:G12*}
G^*_{1,2} = e^{-m\pi} e^{\mp im\pi} 
\end{equation}

In Table 1, the properties of some sub-critical wavenumber cases are collated,  identifying the points in the bottom frame of Fig. \ref{multi}. For the points $P_1$ to $P_{12}$, we have the analytical values of the location in the ($N_\tau, kL_s$)-plane, given along with the modulus of the right-running mode ($|G_1|$), the normalized phase speed and the normalized group velocity in the table. Additionally, three more points are considered in Table 1, with significantly lower wavenumbers as compared to $P_1$- $P_{12}$, and noted as $Q_1$, $Q_2$ and $Q_3$. As the amplification factors varies as $e^{-\nu_l k^2 \tau_s/2}$, so these three points will display amplification factors which will be significantly less attenuated, while $Q_1$ lying on the zero $G_{1,imag}$-contour line (as $P_1$), with $N_\tau = 49$ and $k_0L_s = 0.006411414$. The points $Q_2$ and $Q_3$ have the same value of $N_\tau$, while $k_0 L_s$ values less than half and twice that for $Q_1$, respectively. 

\begin{table}
    \begin{center}
    \def~{\hphantom{0}}
    \begin{tabular}{|c|c|c|c|c|c|c|c|}
        \hline
        Sl. no. & $k_0L_s$ & $N_\tau$ & $G_{1, imag}$ & $|G_1|$ & $C_{ph,1}/c$ & $V_{g,1}/c$ \\[3pt]
        \hline
        $P_1-P_{12}$ & $y^m_*$ & $2m\pi/y^m_*$ & 0.0 & $e^{-m\pi}$ & $1/\sqrt{2}$ & $\sqrt{2}$ \\
         \hline
        $Q_1$ & 0.006411414 & 49 & 0.0 & 0.9987 &0.9999915 & 0.9999740   \\
         \hline
        $Q_2$ & 0.003106500 & 49  & -0.15 &0.9996&0.9999980  & 0.9999940   \\
         \hline
        $Q_3$ & 0.012822824 & 49 & -0.58  & 0.9948  &0.9999660  & 0.9999001  \\ 
         \hline
    \end{tabular}
    \caption{Properties for special points shown in the bottom frame of Fig. \ref{multi}, as denoted by ($P_1,~ P_2,~ ..., P_{12}$), and ($Q_1$, $Q_2$, $Q_3$). In the table, $y^m_* = 1.111319$, for the points $P_1$ to $P_{12}$. }
    \label{tab:prop_table}
    \end{center}
\end{table}
\section{Pulse propagation for sub-critical points with phase coherence property}
In Fig. \ref{multi}, sub-critical points with specific phase shift property of amplification factors are displayed by $P_1, P_2, P_3, .... ,P_{12}$, with imaginary parts of $G_{1,2}$ as zero, with the phase shift given by Eq. \eqref{Eq:beta1}. The location of these points are given by Eq. \eqref{Eq:x*y*}, and the properties are given by Eq. \eqref{Eq:cph} in Table 1. To explain these properties, the propagation of a Gaussian pressure pulse in 1D for a medium with $\nu_l = 0.1443 m^2/s$ is considered, with governing equation given by Eq.  \eqref{Eq:13}. 

\subsection{Bromwich contour integral method for Gaussian pulse propagation}

The Bromwich contour integral method \cite{VanderPol_Bremmer, POF1994, BCIM, Sengupta21}, has been used here to solve Eq. \eqref{Eq:13} with the initial condition given in,
\begin{equation}\label{Eq:initial}
    p'_0(x,0)    = e^{(-\alpha(x-x_0)^2)}sin(k_0(x-x_0))
\end{equation}  
Theoretical details of Bromwich contour integral methods can be obtained in the literature \cite{VanderPol_Bremmer, Bers, IFTT, Sengupta21}. Application of this method is provided in detail for spatiotemporal dynamics \cite{POF1994}, in the context of flow instability, and the signal problem for the spatial instability \cite{Gaster}. A brief recap of the method for the present work is given in the following.

The Bromwich contour is taken along the real wavenumber axis with 9080 equidistant points. This corresponds to a physical domain spanning over $-3m \le x \le 3m$, so that the resolved maximum wavenumber is 4754.27$m^{-1}$. The amplitude of the pulse in the spectral plane at an advanced time ($t + \Delta t$) is given by, 
$$\hat{p} (k, t + \tau_s) = \hat{p} (k, t) (G_1 + G_2)/2$$
\noindent where $G_1$ and $G_2$ are as given in Eq. \eqref{Eq:G12_dim}. Having obtained $\hat{p} (k, t + \tau_s)$ along the Bromwich contour, one performs the inverse Fourier transform to obtain the pulse in the physical plane as, $p'(x,t+\tau_s)$. The choice of $\Delta t$ is made smaller than corresponding $\tau_s$, so that the waves are adequately resolved. In the present exercise, the time steps are so chosen that one has more than ten points in one time period. 

\subsection{Results for some typical pulse propagation cases}

The choice of the Gaussian pulse as the initial condition is centered around those discrete points for which $k_0 L_s = 1.111319303$. These are shown here for the points $P_1, P_3 \;{\rm and}\; P_6$ for which $k_0 = 3362.69211531m^{-1}$ for the choice of $L_s = 3.30485 \times 10^{-4}$m. For these sub-critical wavenumber cases, the Bromwich contour is chosen along the real axis of the complex $k$-plane, while choosing $k_{max} = 4753.22m^{-1}$, a value which is adequately large to correctly obtain the response that excites a large range of wavenumbers.

From Eq. \eqref{Eq:x*y*}, one gets the corresponding values of $N_\tau$. For example, $P_1$ is for $m=1$, which in turn provides the value as, $N_\tau = 3.9958188$; for $P_3$ with $m=3$, one gets $N_\tau = 11.987456$ and for $P_6$ with $m=6$, one gets $N_\tau = 23.974912$. The relevance of these values of $N_\tau$ is reflected in the chosen time-scale, $\tau_s$ calculated using $c = 343.114{\rm m/s}$ from $N_\tau$. For the calculated $\tau_s$, one can obtain the corresponding frequencies as:
259825.02Hz for $P_1$; 86615.82Hz for $P_3$; 43304.29Hz for $P_6$. Another point marked in Fig. \ref{multi} as $Q_1$ provides this frequency as 21188.03Hz, which lies in the same zero-contour as is $P_1$.

\begin{figure*}
\centering
\includegraphics[width=0.9\textwidth]{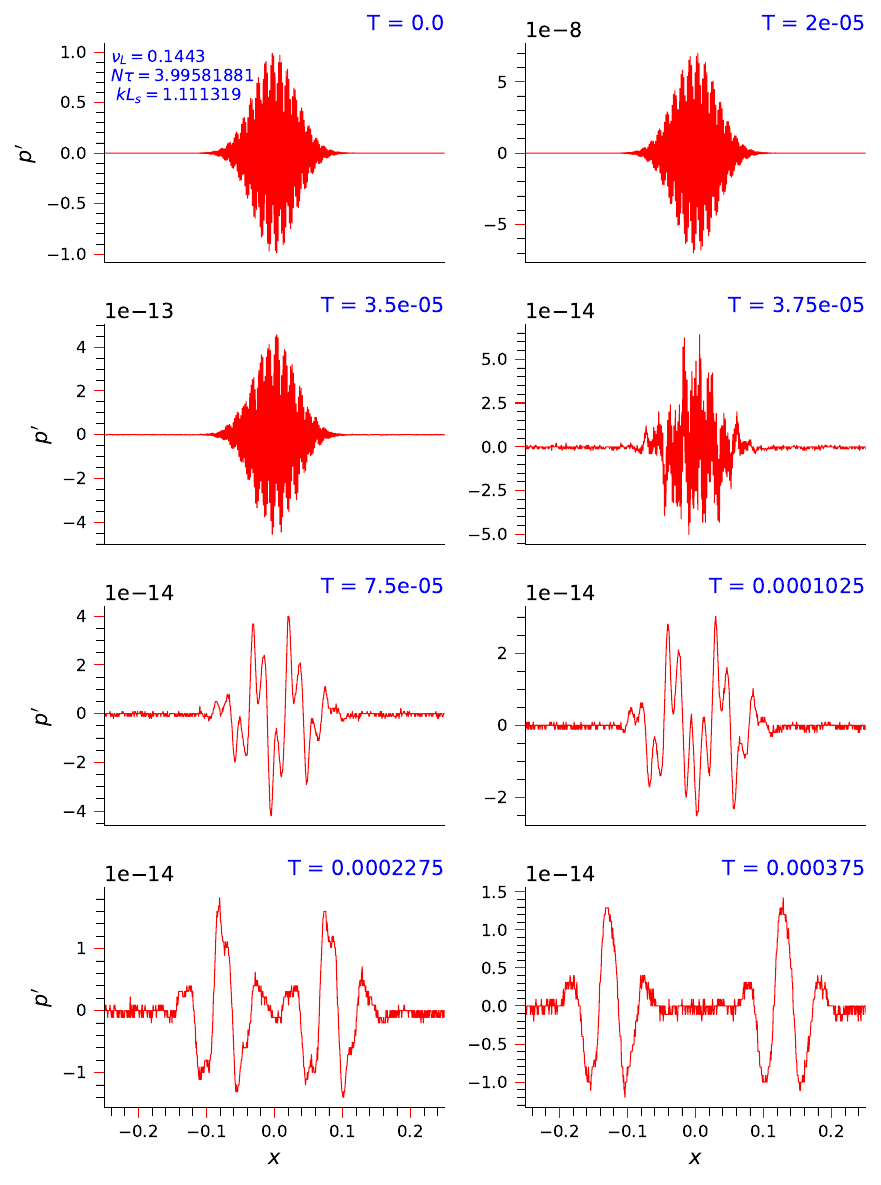}
\caption{The frames show the evolution of the wave-packet for the point $P_1$ marked in Fig. \ref{multi} at indicated time instants.}
\label{P1}
\end{figure*}

In Fig. \ref{P1}, the evolution of the initial wave-packet given by Eq.~\eqref{Eq:initial} for the parameters at $P_1$ is traced for which one will observe the dissipative solution at the indicated times. The attenuation of the initial pulse follows Eq.~ \eqref{Eq:G12_dim} with phase shift corresponding to $\pi$, and is significantly drastic as the rate of attenuation is proportional to the square of the wavenumber in the exponential. As a consequence, the higher wavenumber components decay faster, making the packet smoother at the latter instants shown in various frames. In the final frame at $t=0.000375$s, one can clearly discern the splitting of the initial Gaussian pulse into the left-running and right-running pulses, and the scale shows that all the wavenumber components of the pulse disperse with respect to each other also. 

\begin{figure*}
\centering
\includegraphics[width=0.9\textwidth]{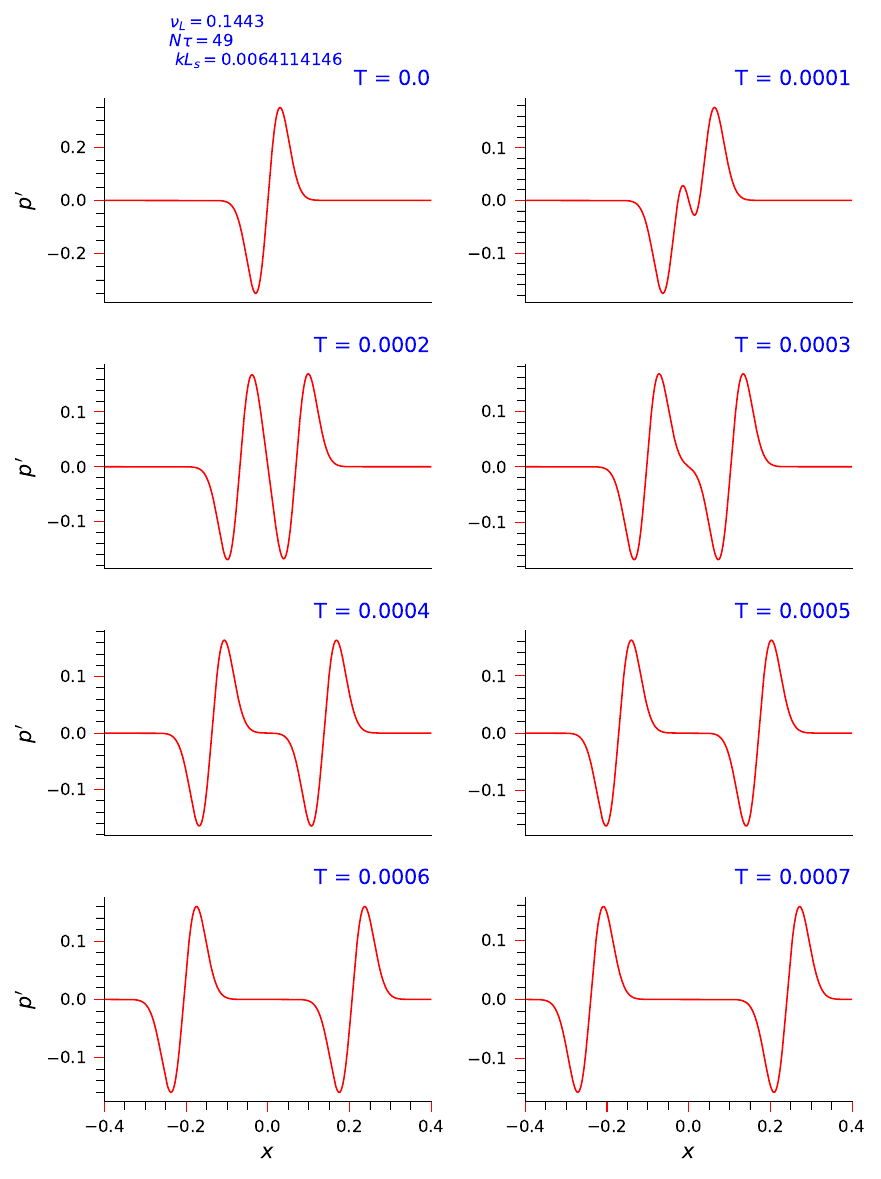}
\caption{The frames show the evolution of the wave-packet for the point $P_3$ marked in Fig. \ref{multi} at indicated time instants.}
\label{P3}
\end{figure*}

\begin{figure*}
\centering
\includegraphics[width=0.9\textwidth]{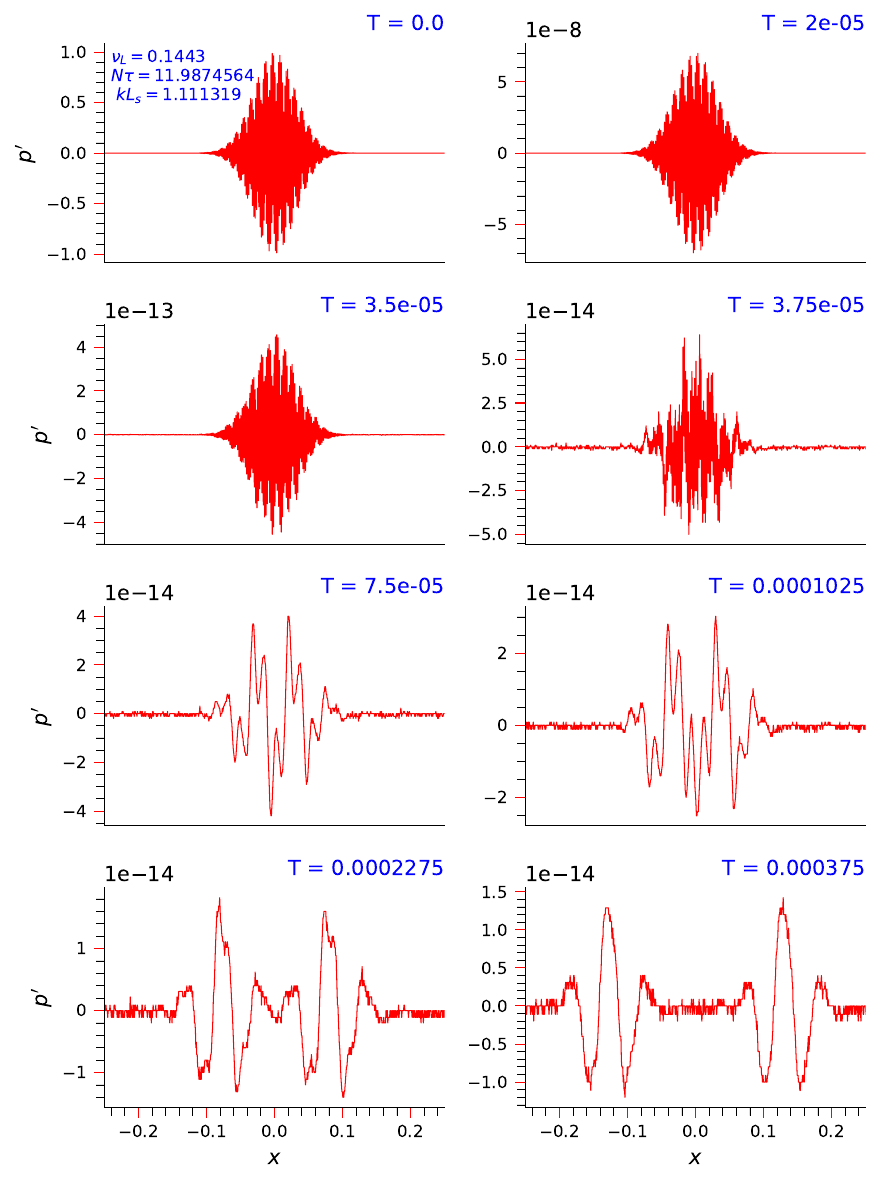}
\caption{The frames show the evolution of the wave-packet for the point $P_6$ marked in Fig. \ref{multi} at indicated time instants.}
\label{P6}
\end{figure*}

In Figs. \ref{P3} and \ref{P6}, the same pulse propagation problem is solved, as it has been shown for Fig. \ref{P1}, i.e. the initial Gaussian pulse given in equation  \eqref{Eq:initial} is used for the same $\alpha =500$ and $k_0 = 3362.69211531m^{-1}$. However, the values of $N_\tau$ for the point $P_3$ and $P_6$ are respectively taken as 11.987456 and 23.974912, implying different imposed time scales given by the corresponding value of $\tau_s$. As shown analytically in the previous section that the phase will be invariant for these discrete points, $P_1$ to $P_{12}$, in Figs. \ref{P1}, \ref{P3} and \ref{P6} the displayed numerical solutions are by Bromwich contour integral method  \cite{POF1994, BCIM, Gaster} for the propagation problem. The imposed time scales are as given by the input pulse for the wavenumbers and the circular frequency values are as obtained by the dispersion relation.

\begin{figure}[H]
\centering
\includegraphics[width=0.9\textwidth]{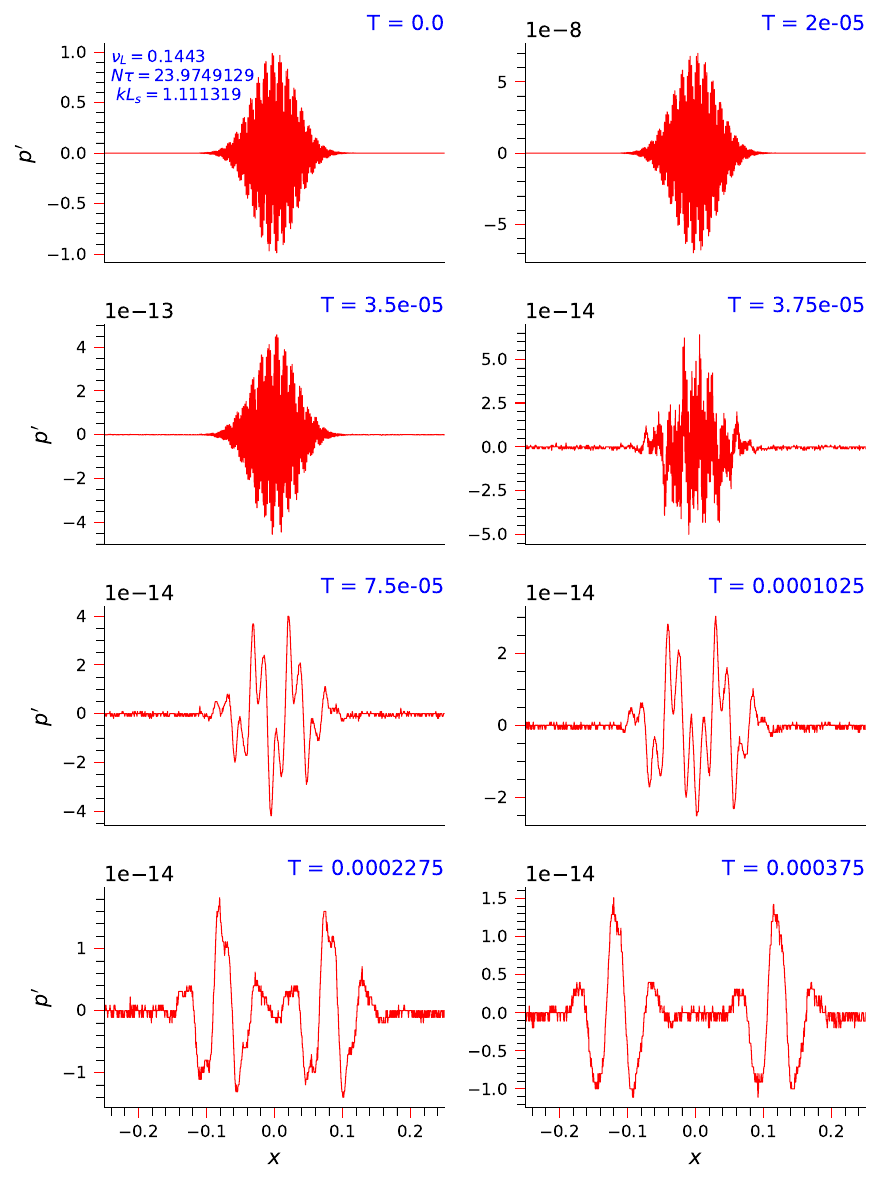}
\caption{The frames show the evolution of the wave-packet for the point $Q_1$ marked in \ref{multi} at different instants of time.}
\label{Q1}
\end{figure}

From Figs. \ref{P1}, \ref{P3} and \ref{P6}, one notices very high attenuation due to the high central value of the wavenumber ($k_0 = 3362.69211531m^{-1})$ of the input Gaussian pulse (appearing as the negative value of the square of the wavenumber) in the exponent of the amplitude. As a consequence, the pulse disappears after a very short time for such high wavenumbers. A similar pulsation and decay of high frequency and high wavenumber pressure fluctuations were noted by \cite{Acoustic_POF13} at very early instants in the RTI. For lower wavenumbers and frequencies, the situation will be qualitatively different for pulse propagation. To investigate such cases, here a simulation is performed for the point, $Q_1$ shown in Fig. \ref{multi}, for which $k_0 = 19.40001698m^{-1}$ (so that $k_0 L_s = 0.0064114146$) and $N_\tau = 49$. Note that this point also belongs to the same zero-contour that passes through $P_1$. 
The corresponding pulse propagation snapshots are shown in Fig. \ref{Q1} for the indicated times up to $t = 0.0007$s, a time for which the amplitude of the pulse remains of the same order, as it is in the initial stage. Various frames of the figure show with clarity the dispersion, along with the splitting of the left-running and right-running wave-packets, as expected for the two modes given by $\omega_{1,2}$.

\begin{figure}[H]
\centering
\includegraphics[width=0.9\textwidth]{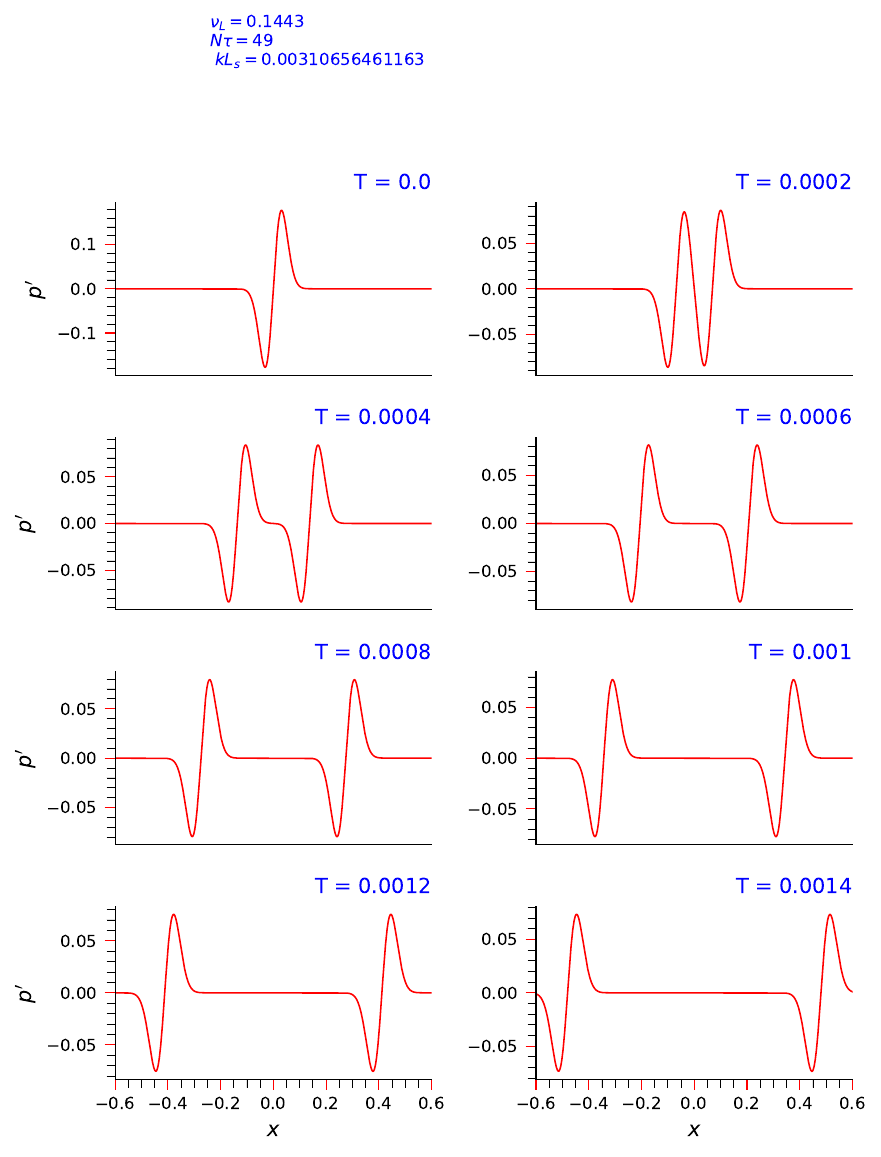}
\caption{The frames show the evolution of the wave-packet for the point $Q_2$ referred in the table\ref{tab:prop_table}.}
\label{Q2}
\end{figure}

\begin{figure}[H]
\centering
\includegraphics[width=0.9\textwidth]{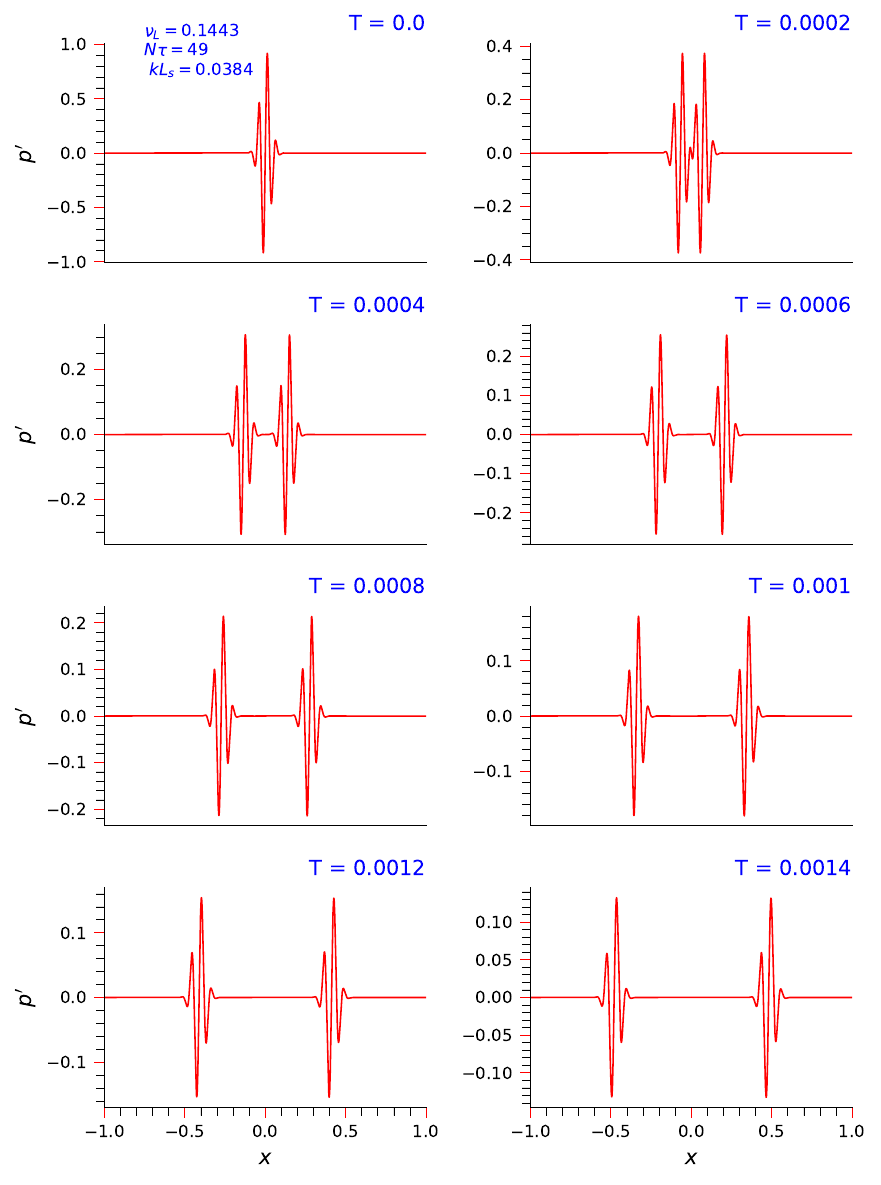}
\caption{The frames show the evolution of the wave-packet for the point $Q_3$ referred in the table\ref{tab:prop_table}}
\label{Q3}
\end{figure}

Additionally, two more cases listed in Table 1 are investigated (marked as Q2 and Q3), with results in Figs. \ref{Q2} and  \ref{Q3} depicting the evolution of the wavepacket at the indicated different time instants. The value of $k_0 L_s$ is $0.0031065006$ for Q2 and $0.012822824$ for Q3. The value of $N_\tau$ is the same as that of Q1 i.e. $N_\tau = 49.0$. It is evident from Fig. \ref{Q2} that the pulse sustains for a longer time as the value of $k_0 L_s$ is the least among all the cases listed in Table 1. The amplitude of the pulse remains of the same order up to $t = 0.0014$, and could be simulated for significantly longer time duration. The wavepacket is observed to split into left-running and right-running wave-packets as expected with the amplitudes of the packets being half of the value noted at $t = 0$. As we are simulating the cases of propagation of the Gaussian wave-packet initially, the lower value of $k_0$ makes the initial peak away from $x=0$, and as a consequence the initial amplitude is less than 1, as given in the initial conditions.

In contrast, the amplitude of the case Q3 undergoes relatively higher attenuation during propagation, as compared to the case of Q2. The separation of the initial pulse into the left-running and right-running packets is also evident, as noted in Fig. \ref{Q3}. The propagation properties of the Gaussian wave-packets are as noted in Table 1.

\newpage

\section{Summary and Conclusions}
The presented research here is a continuation of the study reported earlier \cite{Senguptaetal2024} for flow and acoustic disturbance propagation as fluctuating pressure in a quiescent ambience. The linearized compressible Navier-Stokes equation for the disturbance field is studied because the signal has associated fluctuating pressure which is ten to eleven orders of magnitude smaller than the hydrodynamic pressure. Analysis of the governing equation in the spectral plane with the wavenumber and circular frequency as independent variables yields the quadratic dispersion relation, as introduced earlier \cite{Senguptaetal2023}. Compared to the investigation in the physical plane reported \cite{Senguptaetal2024}, here the adoption of Bromwich contour integral allows us to obtain higher accuracy solution for the signal problem.  

Contrary to the classical wave equation, the role of viscous stresses makes the studied system both dissipative and dispersive. The viscous loss terms were included for planar propagation in the literature \cite{Blackstock2000, Trusler, Morse_Ingard}, but comprehensive analyses are communicated recently \cite{Senguptaetal2024} and here.   

Obtained properties of disturbance propagation are furthermore verified by numerically computing pulse propagation by Bromwich contour integral method \cite{VanderPol_Bremmer, IFTT, POF1994, Gaster, BCIM}. This aspect of signal propagation for pulses has been reported here for the first time. 


An interesting aspect of GSA applied to the pulse propagation is reported when the phase shift between the real and imaginary parts of the amplification factors becomes integral multiple of $\pi$. This manifests in the disappearance of the imaginary part of $G_1$ and $G_2$, as given by Eq. \eqref{Eq:locus}. Such discrete solutions can have infinite occurrences. In Fig. \ref{multi}, this is demonstrated by plotting the imaginary parts of $G_1$ and $G_2$ in the top two frames, which shows many such zero-contours. In the bottom frame of this figure, twelve such zero-contours are plotted and the apex of these contours are identified as $P_1$, $P_2$, $P_3$ ..... $P_{12}$ which identify the least time scales for each of these cases with integer $m$'s. It is to be emphasized that even though $G_1$ and $G_2$ are strictly real, these sub-critical cases represent attenuated waves, and not as strictly diffusive case noted for $k =k_c$, i.e. $m=0$. The values of $N_\tau$ and $k_mL_s$ are identified in Eq.~\eqref{Eq:locus} for these special points. It has been established that the phase speed of these least-$\tau_s$ discrete modes is equal to $c/\sqrt{2}$, and the group velocity is equal to $c\sqrt{2}$ for $P_1$ to $P_{12}$.

The propagation of a Gaussian pulse for these special points with identical $k_0L_s$ is reported in Figs. \ref{P1}, \ref{P3} and \ref{P6}. The evanescent pulses for these cases indicate their short life span before attenuation below machine zero, for the numerical solution obtained by Bromwich contour integral method \cite{VanderPol_Bremmer, IFTT, POF1994}. Such pressure pulse propagation is very vitally important in triggering RTI, as noted earlier \cite{Acoustic_POF13} from the interface. The figures clearly show identical propagation properties and indicate a clear separation of the left-running and right-running waves. The importance of the present analysis and the Bromwich contour integral for pulse propagation is established by considering another case for the point $Q_1$ in Fig. \ref{multi}, which belongs to the same zero-contour as $P_1$. For the significantly lower value of $k_0$ for this case, the pulse attenuates at a much lower rate, as noted in Fig. \ref{Q1}. In this case, the group velocity is virtually equal to $c$. The difference of the energy propagation group velocity for $P_1$ and $Q_1$ clearly demonstrate the dispersive nature of the pulse propagation due to the added viscous diffusion term in the analysis.

Additionally, two more cases have been studied for pulse propagation, which are listed as Q2 and Q3 in Table 1, having the same value of $N_\tau$ as is the case for Q1. While Q2 has the central wavenumber of the initial Gaussian wave-packet which is half of the value that is considered for Q1, the central wavenumber for Q3 is double the value of that is considered for Q1. Based on these wavenumber amplitudes, the case of Q2 shows the least attenuation, while Q1 has the intermediate attenuation, while Q3 has the highest attenuation among these three cases. 

Having established the correctness of the novel analysis of pulse propagation in quiescent ambience, and the robustness of the Bromwich contour integral method, it will be implemented for newer developments, including those cases where there is convection in the ambience.

\section{Acknowledgement}
The authors acknowledge various discussions with Prof. Aditi Sengupta (IIT Dhanbad) and Prof. K. R. Sreenivasan (NYU). 

\section*{AUTHOR DECLARATIONS}
\subsection*{Conflict of Interest}
The authors have no conflicts to disclose.

\section*{DATA AVAILABILITY}
The data that support the findings of this study are available from the corresponding author upon reasonable request.

\bibliography{acoustics_edited}
\end{document}